\documentclass{ws-procs975x65}

\def\beq{\begin{equation}}
\def\eeq{\end{equation}}

\begin{document}

\def \d {{\rm d}}

\def \bm #1 {\mbox{\boldmath{$m_{(#1)}$}}}

\def \bF {\mbox{\boldmath{$F$}}}
\def \bV {\mbox{\boldmath{$V$}}}
\def \bff {\mbox{\boldmath{$f$}}}
\def \bT {\mbox{\boldmath{$T$}}}
\def \bk {\mbox{\boldmath{$k$}}}
\def \bl {\mbox{\boldmath{$\ell$}}}
\def \bn {\mbox{\boldmath{$n$}}}
\def \bbm {\mbox{\boldmath{$m$}}}
\def \tbbm {\mbox{\boldmath{$\bar m$}}}

\def \0 {^{(0)}}
\def \1 {^{(1)}}

\def \a {\alpha}
\def \B {\beta}

\def \T {\bigtriangleup}
\newcommand{\msub}[2]{m^{(#1)}_{#2}}
\newcommand{\msup}[2]{m_{(#1)}^{#2}}

\newcommand{\be}{\begin{equation}}
\newcommand{\ee}{\end{equation}}

\newcommand{\beqn}{\begin{eqnarray}}
\newcommand{\eeqn}{\end{eqnarray}}
\newcommand{\AdS}{anti--de~Sitter }
\newcommand{\AAdS}{\mbox{(anti--)}de~Sitter }
\newcommand{\AAN}{\mbox{(anti--)}Nariai }
\newcommand{\AS}{Aichelburg-Sexl }
\newcommand{\pa}{\partial}
\newcommand{\pp}{{\it pp\,}-}
\newcommand{\ba}{\begin{array}}
\newcommand{\ea}{\end{array}}

\newcommand{\M}[3] {{\stackrel{#1}{M}}_{{#2}{#3}}}
\newcommand{\m}[3] {{\stackrel{\hspace{.3cm}#1}{m}}_{\!{#2}{#3}}\,}

\newcommand{\tr}{\textcolor{red}}
\newcommand{\tb}{\textcolor{blue}}
\newcommand{\tg}{\textcolor{green}}
\newcommand{\tor}{\textcolor{orange}}

\def\a{\alpha}
\def\g{\gamma}
\def\de{\delta}

\def\b{{\kappa_0}}

\def\R{{\cal R}}
\def\F{{\cal F}}
\def\L{{\cal L}}

\def\e{e}
\def\bb{b}

\title{VSI electromagnetic fields}

\author{Marcello Ortaggio and Vojt\v ech Pravda} 
\address{Institute of Mathematics of the Czech Academy of Sciences \\ \v Zitn\' a 25, 115 67 Prague 1, Czech Republic \\
$^*$E-mail: ortaggio(at)math(dot)cas(dot)cz, pravda@math.cas.cz}

\begin{abstract}
	A $p$-form $\bF$ is VSI (i.e., all its scalar invariants of arbitrary order vanish) in a $n$-dimensional spacetime {\em if and only if} it is of type N, its multiple null direction $\bl$ is ``degenerate Kundt'', and {$\pounds_{\bl}\bF=0$}. This recent result is reviewed in the present contribution and its main consequences are summarized. In particular, a subset of VSI Maxwell fields possesses a {\em universal} property, i.e., they also solve (virtually) any generalized (non-linear and with higher derivatives) electrodynamics, possibly also coupled to Einstein's gravity.
\end{abstract}

\bodymatter

\section{Introduction}

\label{sec_intro}

The present contribution is summary of the main results of our recent work \cite{OrtPra15}. 
It is useful to start by defining the VSI property for a general tensor, i.e.,
\begin{definition}[VSI tensors]
	\label{def_VSI}
 A tensor in an $n$-dimensional spacetime with metric $g_{ab}$ is VSI$_I$ if all the scalar polynomial invariants constructed from the tensor itself and its covariant derivatives up to order $I$ ($I=0,1,2,3,\dots$) vanish. It is VSI if all its scalar polynomial invariants of {\em arbitrary order} vanish. 
\end{definition}

As a generalization of the notion of null fields \cite{syngespec} to arbitrary $p$-forms, it is natural to introduce the following 
\begin{definition}[$p$-forms of type N]
	\label{def_N}
 At a spacetime point, a $p$-form $\bF$ is of type~N if it satisfies
\be
	\ell^a F_{a b_1\ldots b_{p-1}}=0 , \qquad \ell_{[a}F_{b_1\ldots b_{p}]}=0 ,
	\label{BelDeb}
\ee
where $\bl$ is a null vector (this follows from \eqref{BelDeb} and need not be assumed). The second condition can be equivalently replaced by $\ell^a\, {}^*F_{a b_1\ldots b_{n-p-1}}=0$.
\end{definition}
This is equivalent to the type N condition in the set-up of \cite{Milsonetal05}. 

First, with the results of\cite{Hervik11}, it is easy to see that {\em a $p$-form $\bF$ is VSI$_0$ iff it is is of type N}. Next, our main result is the VSI condition, given in the next section.

\section{Main result: VSI $p$-forms}

The main result of\cite{OrtPra15} is the following
\begin{theorem}[VSI $p$-forms\cite{OrtPra15}]
 \label{theor}
The following two conditions are equivalent:

\begin{enumerate}

\item\label{cond1} a non-zero $p$-form field $\bF$ is VSI in a spacetime with metric $g_{ab}$
\item\label{cond2} 
 \begin{enumerate}
		\item\label{cond2a} $\bF$ possesses a multiple null direction $\bl$, i.e., it is of type N
		\item\label{cond2b} {$\pounds_{\bl}\bF=0$} 
		\item\label{cond2c} $g_{ab}$ is a {\em degenerate Kundt} metric, and $\bl$ is the corresponding Kundt null direction.
 \end{enumerate}

\end{enumerate}

\end{theorem}

We observe that, in Theorem~\ref{theor}, the $p$-form $\bF$ is not assumed to satisfy any particular field equations and the result is thus purely geometric (on the other hand, {\em if} $\bF$ is taken to be  {\em closed}, i.e., $\d\bF=0$, then condition~\ref{cond2b} automatically follows from the type N condition \ref{cond2a}, and need not be assumed).

\subsection{Adapted coordinates}

From Theorem~\ref{theor} with the results of \cite{ColHerPel09a,Coleyetal09}, it follows that coordinates   $(u,r,x^\alpha)$, adapted to $\bl=\partial_r$, exist such that any VSI $p$-form can be written as
\be
 \bF=\frac{1}{(p-1)!}f_{\alpha_1\ldots\alpha_{p-1}}(u,x)\d u\wedge\d x^{\alpha_1}\wedge\ldots\wedge\d x^{\alpha_{p-1}} ,
 \label{F_N_coords}
\ee
and the corresponding background metric as
\beqn
 & & \d s^2 =2\d u\left[\d r+H(u,r,x)\d u+W_\alpha(u,r,x)\d x^\alpha\right]+ g_{\alpha\beta}(u,x) \d x^\alpha\d x^\beta , \label{Kundt_gen} \\
 & & W_{\alpha}(u,r,x)=rW_{\alpha}^{(1)}(u,x)+W_{\alpha}^{(0)}(u,x) , \label{deg_Kundt1} \\
 & & H(u,r,x)=r^2H^{(2)}(u,x)+rH^{(1)}(u,x)+H^{(0)}(u,x) , \label{deg_Kundt2}
\eeqn
where $\alpha,\beta=2 \dots n-1$, and $x$ denotes collectively the set of coordinates $x^\alpha$.

The degenerate Kundt metric~\eqref{Kundt_gen}--\eqref{deg_Kundt2} includes, e.g., all VSI spacetimes\cite{Pravdaetal02,Coleyetal04vsi}, all \pp waves, and all Kundt Einstein (or aligned pure radiation) spacetimes -- in particular, spacetimes of constant curvature (Minkowski and (A)dS). The corresponding Weyl and Ricci tensors are both of aligned type II, in general.

If $\bF$ in \eqref{F_N_coords} is required to obey the source-free Maxwell equations $\d\bF=0=\d^*\bF$, the following conditions are obtained
\be
   f_{[\a_2\ldots\a_{p-1},\a_1]}=0 , \qquad  (\sqrt{\tilde g}\,f^{\beta\a_1\ldots\a_{p-2}})_{,\beta}=0 , 
	\label{Maxwell}
\ee
where $\tilde g\equiv\det g_{\alpha\beta}=-\det g_{ab}\equiv -g$. These are Maxwell's equations for the $(p-1)$-form $\bff$ in the $(n-2)$-dimensional Riemannian geometry associated with $g_{\alpha\beta}$, i.e., $\bff$ must be {\em harmonic} (recall, however, that $\bff$ can also depend on $u$). For $n=3,4$, it can be argued that to any $\bF$ which is VSI, it can always be associated a solution of the Maxwell equations which is also VSI\cite{OrtPra15}. It is also interesting to observe that the effective Maxwell equations \eqref{Maxwell} are ``immune'' to adding a Chern-Simons term (except when this is linear, see\cite{OrtPra15} for details) to the full Maxwell equations.

\subsection{Universal solutions of generalized electrodynamics}

\label{subsec_univ_test}

It was already known to Schr{\"o}dinger \cite{Schroedinger35,Schroedinger43} that all null Maxwell fields (with $n=4=2p$) solve the equations for the electromagnetic field in any non-linear electrodynamics. More generally, we observe that a subset of the VSI Maxwell fields described above possesses a ``universal'' property, i.e., they solve simultaneously any electrodynamics whose field equations can be expressed as $\d\bF=0$, $*\d\!*\!\!\tilde\bF=0$, where $\tilde\bF$ can be any $p$-form constructed from $\bF$ and its covariant derivatives. For example, any VSI Maxwell $\bF$ is universal if the background is a Kundt spacetime of Weyl and traceless-Ricci type III (aligned) with $DR=0=\delta_i R$ (an affine parameter and a frame parallelly transported along $\bl$ are assumed -- cf.\cite{OrtPra15} for the notation employed here). In particular, Ricci flat and Einstein Kundt spacetimes of Weyl type III/N/O can occur, the latter including Minkowski and (A)dS. 

An explicit simple example is given by the Maxwell field
\be
	\bF=e^{x/2}c(u)\d u\wedge\left(-\cos\frac{ye^u}{2}\d x+e^u\sin\frac{ye^u}{2}\d y\right) ,
	\label{F_example}
\ee
defined in the type III vacuum spacetime found by Petrov (eq.~(31.40) in \cite{Stephanibook})
\be
 \d s^2 =2\d u\left[\d r+\frac{1}{2}\left(xr-xe^x\right)\d u\right]+ e^x(\d x^2+e^{2u}\d y^2) . \label{Petrov} \\
\ee

More general results will be presented elsewhere.

\section{Einstein-Maxwell solutions}

\label{sec_EM}

\subsection{General field equations}

The previous discussion applies to VSI test fields, since we have not considered the consequences of the backreaction on the spacetime geometry. In the full Einstein-Maxwell theory with an arbitrary cosmological constant $\Lambda$, one finds that
the metric functions entering \eqref{Kundt_gen} must satisfy the following set of equations (obtained in\cite{OrtPra15} refining the results of\cite{PodZof09}) 
\beqn
 & & \R_{\alpha\beta}=\frac{2\Lambda}{n-2}g_{\alpha\beta}+\frac{1}{2}W_{\alpha}^{(1)}W_{\beta}^{(1)}-W_{(\alpha||\beta)}^{(1)} , \label{Rij} \\
 & & 2H^{(2)}=\frac{\R}{2}-\frac{n-4}{n-2}\Lambda+\frac{1}{4}W^{(1)\alpha}W_{\alpha}^{(1)} , \label{H2} \\
 & & W_{\a||\B}^{(1)\ \B}=\frac{1}{2}W^{(1)\B}\left(3W_{\a||\B}^{(1)}-W_{\B||\a}^{(1)}\right)+W^{(1)}_{\a}\left(W^{(1)\B}_{\ \ \ \ \ ||\B}-\frac{1}{2}W^{(1)\B}W^{(1)}_\B -\frac{2 \Lambda}{n-2}\right) , \\
 & & 2H^{(1)}_{,\a}=-{g_{\a\B,u}}^{||\B}+2W^{(0)\ \ \ \B}_{[\a||\B]}-2W^{(0)\B}W^{(1)}_{\a||\B}+(W^{(0)\B}W^{(1)}_\B)_{,\a}+W^{(1)}_{\a,u}+2(\ln\sqrt{\tilde g})_{,u\a} \nonumber \\
 & & \quad {}+W^{(1)}_\a\left[W^{(0)\B}W^{(1)}_\B-W^{(0)\B}_{\ \ \ \ \ ||\B}+(\ln\sqrt{\tilde g})_{,u}\right]+\frac{4\Lambda}{n-2}W^{(0)}_\a , \label{H1} \\
 & & \Delta H^{(0)}+W^{(1)\a}H^{(0)}_{,\a}+W^{(1)\a}_{\ \ \ \ \ ||\a}H^{(0)}=W^{(0)\B}W^{(0)}_\B\left(\frac{1}{2}W^{(1)\a}_{\ \ \ \ \ ||\a}-\frac{2\Lambda}{n-2}\right) \nonumber \\
 & & \quad {}+H^{(1)}\left[W^{(0)\a}_{\ \ \ \ \ ||\a}-(\ln\sqrt{\tilde g})_{,u}\right]-\frac{1}{2}(W^{(0)\a}W^{(1)}_\a)^2+W^{(0)[\a||\B]}W^{(0)}_{[\a||\B]}+W^{(0)\ ||\a}_{\a,u} \nonumber \\
 & & \quad {}-W^{(0)\B}\left(2W^{(1)\a}W^{(0)}_{[\a||\B]}+W^{(1)}_{\B,u}-2H^{(1)}_{,\B}\right)-(\ln\sqrt{\tilde g})_{,uu}+\frac{1}{4}g^{\a\B}_{\ \ ,u}g_{\a\B,u} -\b\F^2 . \label{H0}
\eeqn
Heree $\R_{\alpha\beta}$, $\R$ and $||$ denote, respectively, the Ricci tensor, the Ricci scalar and the covariant derivative associated with $g_{\alpha\beta}$, $W^{(1)\alpha}\equiv g^{\alpha\beta}W_{\beta}^{(1)}$, $\Delta$ is the Laplace operator in the geometry of the transverse metric $g_{\a\B}$, and $\b$ is a gravitational coupling constant. The Maxwell equations~\eqref{Maxwell} must also be satisfied.

The simplest examples one can construct are electromagnetic and gravitational ``plane-fronted'' waves (with $W_{\alpha}^{(0)}=0$) propagating in a constant curvature background, giving rise to Kundt waves of Weyl type N (in four dimensions see, e.g., \cite{GarPle81,OzsRobRoz85}). More general (e.g., with $W_{\alpha}^{(0)}\neq0$) degenerate Kundt metrics with null Maxwell fields are also known (see \cite{Stephanibook,GriPodbook,GriDocPod04} and references therein for $n=4$). The case of VSI $p$-form Maxwell fields in VSI and \pp waves spacetimes have been discussed in\cite{OrtPra15} (where further references can be found).

\subsection{Universal Einstein-Maxwell solutions}

\label{subsec_univ_EM}

Some of the universal Maxwell fields mentioned in section~\ref{subsec_univ_test} can also be used to construct exact solutions of full general relativity, where the energy-momentum tensor $T_{ab}$ associated with the electromagnetic field is determined in the generalized electrodynamics (in terms of $\bF$ and its covariant derivatives -- cf. \cite{Podolsky42} for an example). For example, as pointed out in\cite{OrtPra15}, all VSI spacetimes with $L_{i1}=0=L_{1i}$ (i.e., the recurrent ones) coupled to an aligned VSI $p$-form field that solve the standard Einstein-Maxwell equations are also exact solutions of gravity coupled to generalized electrodynamics, provided $p>1$ and $\delta_iF_{1j_1\ldots j_{p-1}}=0$ (in an ``adapted'' parallely transported frame, i.e., such that $\M{i}{j}{k}=0$). Within this family, metrics of Weyl type N are necessarily \pp waves, for which such a universal property {was pointed out in} \cite{Guven87,HorSte90,Horowitz90}, at least for certain values of $p$. But metrics of Weyl type III are also permitted, including \pp waves ($L_{11}=0$) and also genuinely recurrent ($L_{11}\neq0$) spacetimes (for $n=4$, $p=3$ this was discussed in \cite{Coley02}). One explicit example of the latter solutions in 4D is given by the Maxwell field~\eqref{F_example} with the metric
\be
 \d s^2 =2\d u\left[\d r+\frac{1}{2}\left(xr-xe^x-2\b e^xc^2(u)\right)\d u\right]+ e^x(\d x^2+e^{2u}\d y^2) , \label{Petrov_EM} 
\ee
which is a modification of \eqref{Petrov} taking into account the backreaction.

As in section~\ref{subsec_univ_test}, the above discussion applies to generalized electrodynamics with arbitrary higher-order derivative ``corrections''. A special instance of this result is the fact that Einstein-Maxwell solutions with aligned null electromagnetic fields (not necessarily VSI) are also solution of NLE coupled to gravity, as previously demonstrated in \cite{Kichenassamy59,KreKic60,Peres61}.

\section{Further remarks}

For certain purposes, it may be useful to observe that if $\bF$ is VSI$_3$ then it is necessarily VSI (this follows from the proof of Theorem~\ref{theor} given in\cite{OrtPra15}). For completeness, let us thus also give the necessary and sufficient conditions for a $p$-form $\bF$ to be VSI$_1$ or VSI$_2$ (recall that VSI$_0$ means type N, as mentioned in section~\ref{sec_intro}):
\begin{proposition}[VSI$_1$ and VSI$_2$ $p$-forms\cite{OrtPra15}]
 \label{proposition_VSI1_2}
	A $p$-form $\bF$ is VSI$_1$ iff it is is of type N, {$\pounds_{\bl}\bF=0$}, $\bl$ is Kundt. It is VSI$_2$ iff it is is of type N, {$\pounds_{\bl}\bF=0$}, $\bl$ is Kundt and (at least) doubly aligned with the Riemann tensor.
\end{proposition} 

It is also interesting to observe that Theorem~\ref{theor} and Proposition~\ref{proposition_VSI1_2} apply also in the limiting case $p=1$, i.e., when $\bF$ is a vector field. In particular, when $p=1$, Theorem~\ref{theor} reduces to: {\em a vector field $\bl$ is VSI in a spacetime with metric $g_{ab}$ iff $\bl$ is Kundt and affinely parameterized, and $g_{ab}$ is a degenerate Kundt metric w.r.t $\bl$}. 

The above comments are summarized in the first two columns of Fig.~\ref{table}. The last column gives corresponding results in the case of the Riemann tensors (i.e., for the VSI$_I$ spacetimes)\cite{Pravdaetal02,Coleyetal04vsi,Pelavasetal05}. While certain conditions turn out to be similar, there is also an important difference: for the Riemann tensor one has already $VSI_2\Rightarrow VSI$.

\begin{figure}[t]
 \begin{center}
	\begin{tabular}{|c|l|l|l|}
	\hline
			 &&&\\[-3mm]
			 & \ vector $\bl$ \  & \ \ $p$-form $\bF$ \ \  & \ \ Riemann  \ \  \\
			 &&&\\[-3mm]
			 \hline &&&\\[-10pt]
    VSI$_0$	& \hspace{-.25cm} null  & \hspace{-.25cm} N  & \hspace{-.25cm}  III (N,O) \\[1pt]
		\hline &&&\\[-10pt]
		VSI$_1$ & \hspace{-.25cm} Kundt & \hspace{-.25cm} N, $\pounds_{\bl}\bF=0$, Kundt & \hspace{-.25cm} (*) N, \!$\kappa=0$, $\sigma\Psi_4\!=\!\rho\Phi_{22}$ \\[1pt]
		\hline &&&\\[-10pt]
		VSI$_2$ & \hspace{-.25cm} Kundt, Riem II &  \hspace{-.25cm} N, $\pounds_{\bl}\bF=0$, Kundt, Riem II & \hspace{-.25cm} III, Kundt \\[1pt]
		\hline &&&\\[-10pt]
		VSI$_3$ & \hspace{-.25cm} degKundt & \hspace{-.25cm}  N, $\pounds_{\bl}\bF=0$, degKundt & \hspace{-.25cm} `` \\[1pt]
		\hline &&&\\[-10pt]
		\vdots & \hspace{-.25cm} `` \ & \hspace{-.25cm} `` \ & \hspace{-.25cm} `` \\[1pt]
		\hline &&&\\[-10pt]
		VSI & \hspace{-.25cm} `` \ & \hspace{-.25cm} `` \ & \hspace{-.25cm} `` \ \\[1pt]
		\hline 
		\end{tabular}
		\caption{Comparing various VSI$_I$ conditions for a 1-form $\bl$, a $p$-form $\bF$\cite{OrtPra15} and the Riemann tensor\cite{Pravdaetal02,Coleyetal04vsi,Pelavasetal05}. The quotation marks `` mean that the same conditions as in the next-upper row apply. The Riemann VSI$_1$ condition (*) needs some comments for two reasons. First, it has been investigated only in the case $n=4$\cite{Pelavasetal05} (and indeed it is given above in the standard 4D NP notation). Additionally, it is sufficient but not necessary (contrary to the rest of the table): also Kundt spacetimes of Riemann type III are VSI$_1$ (but not ``properly'', i.e., they are in fact VSI, cf. the lower rows in the same column).}
		\label{table}
	\end{center}
\end{figure}



\section*{Acknowledgments}

We thank research plan {RVO: 67985840} and research grant GA\v CR 13-10042S.

\end{document}